\documentclass[12pt,titlepage]{article}
\usepackage{graphicx}
\usepackage{amsmath}
\usepackage{amsfonts}
\usepackage{amssymb}
\usepackage{mychicago}
\usepackage{subfigure}
\usepackage{genetics}
\usepackage{natbib}
\usepackage{multirow}

\newcommand{\Prob}{\operatorname{P}}
\newcommand{\given}{\,|\,}

\newcommand{\COV}{\operatorname{COV}}

\title{Multiple Quantitative Trait Analysis Using Bayesian Networks \\
\protect\small A ``Methods, Technology, and Resources'' Article submitted
  to {\em Genetics}}

\author{Marco Scutari\thanks{Genetics Institute, University College London (UCL), 
          United Kingdom}, 
        Phil Howell\thanks{National Institute of Agricultural Botany (NIAB),
          Cambridge, United Kingdom},
        David J. Balding\footnotemark[1],
        Ian Mackay\footnotemark[2]}

\renewcommand{\CorrespondingAddress}{
  UCL Genetics Institute (UGI) \\
  University College London \\
  Gower Street \\
  London WC1E 6BT \\
  United Kingdom \\
  tel.: +44(0)2067692212 \\
  email: \texttt{m.scutari@ucl.ac.uk} \vfill}
\renewcommand{\RunningHead}{Multiple Trait Analysis With BNs}
\renewcommand{\CorrespondingAuthor}{Marco Scutari}
\renewcommand{\KeyWords}{Multiple Traits, Quantitative Traits, Bayesian Networks,
  Genome-Wide Predictions.}

\begin{document}

\maketitle

\begin{abstract}

Models for genome-wide prediction and association studies usually target a 
single phenotypic trait. However, in animal and plant genetics it is common to
record information on multiple phenotypes for each individual that will be
genotyped. Modeling traits individually disregards the fact that they are
most likely associated due to pleiotropy and shared biological basis, thus
providing only a partial, confounded view of genetic effects and phenotypic
interactions. In this paper we use data from a Multiparent Advanced Generation
Inter-Cross (MAGIC) winter wheat population to explore Bayesian networks as a
convenient and interpretable framework for the simultaneous modeling of multiple
quantitative traits. We show that they are equivalent to multivariate genetic
best linear unbiased prediction (GBLUP), and that they are competitive with
single-trait elastic net and single-trait GBLUP in predictive performance.
Finally, we discuss their relationship with other additive-effects models and
their advantages in inference and interpretation. MAGIC populations provide an
ideal setting for this kind of investigation because the very low population
structure and large sample size result in predictive models with good power
and limited confounding due to relatedness. 

\end{abstract}

\section{Introduction}

Understanding the behavior of complex traits involves modeling a web of
interactions among the effects of genes, environmental conditions and other 
covariates. Ignoring one or more of these factors may substantially impact the
accuracy and the generality of the conclusions that can be drawn from the model
\citep{hartley,eeuwijk,sem}, both in the context of genome-wide association
studies (GWAS) and genomic selection (GS). Indeed a lot of attention has been
devoted in recent literature to improving traditional additive genetic models,
which were originally defined using only allele counts \citep[e.g.][]{meuwissen},
by supplementing them with additional information. Some examples include 
marker-based kinship coefficients \citep{doug}, spatial heterogeneity and
dominance \citep{finley}, and gene expression data \citep{robbie}.

However, most studies in plant and animal genetics still focus on a single 
phenotypic trait at a time despite the availability of a set of simultaneously
measured traits for each genotyped individual. Models for analyzing multiple
traits have been available since \citet{henderson} introduced the multivariate
extension of the genetic best linear unbiased prediction (GBLUP) models, and
have been investigated as recently as \citet{stephens} in the context of GWAS.
More recent additions include structural equation models \citep[SEM;][]{sem},
a Bayesian extension of seemingly unrelated regression \citep[SUR;][]{sur}, the
MultiPhen ordinal regression \citep{paul} and spatial models \citep{banerjee}.

In this paper we will use Bayesian networks \citep[BNs;][]{pearl,koller} to
build a multivariate dependency model that accounts for simultaneous associations
and interactions among multiple single nucleotide polymorphisms (SNPs) and 
phenotypic traits. BNs have been applied to the analysis of several kinds of 
genomic data such as gene expression \citep{friedman}, protein-protein 
interactions \citep{jansen,sachs}, pedigree analysis \citep{lauritzen} and the
integration of heterogeneous genetic data \citep{integrative}. Their modular
nature makes them ideal for analyzing large marker profiles. As far as SNPs are
concerned, BNs have been used to investigate linkage disequilibrium 
\citep[LD;][]{gianola,sinoquet} and epistasis \citep{epibn}, and to determine
disease susceptibility for anemia \citep{anemia}, leukemia \citep{integrative},
and hypertension \citep{malovini}. The same BN can simultaneously highlight SNPs
potentially involved in determining a trait (e.g. for association purposes) and
be used for prediction (e.g. for selection purposes): a network capturing the
relationship between genotypes and phenotypes can be used to compute the 
probability that a new individual with a particular genotype will have the 
phenotype of interest \citep{lauritzen,cowell}.

\section{Materials and Methods}

A Bayesian network (BN) is a probabilistic model in which a directed acyclic
graph $G$ is used to define the stochastic dependencies quantified by a
probability distribution \citep{pearl,koller}. The variables $\mathbf{X} =
\{X_i\}$ under investigation in this context include $T$ traits $X_{t_1},
\ldots, X_{t_T}$ and $S$ SNPs $X_{s_1}, \ldots, X_{s_S}$, each of which is
associated with a node in $G$. The arcs between the nodes represent direct
stochastic dependencies, and determine how the \textit{global distribution} of
$\mathbf{X}$ decomposes into a set of \textit{local distributions},
\begin{equation}
\label{eq:markov}
  \Prob(\mathbf{X}) = \prod \Prob(X_i \given \Pi_{X_i});
\end{equation}
one for each variable $X_i$, depending only on its parents $\Pi_{X_i}$. This 
modular representation can capture direct and indirect associations between
SNPs and phenotypes; and associations between SNPs due to linkage and population
structure.

In the spirit of commonly used additive genetic models for quantitative traits
\citep[e.g.][]{meuwissen}, we make some further assumptions on the BN:
\begin{enumerate}
  \item each variable $X_i$ is normally distributed, and $\mathbf{X}$ is
    multivariate normal; \label{pt1}
  \item stochastic dependencies are assumed to be linear; \label{pt2}
  \item traits can depend on SNPs (i.e. $X_{s_i} \rightarrow X_{t_j}$) but not
    vice versa (i.e. not $X_{t_j} \rightarrow  X_{s_i}$), and they can depend
    on other traits (i.e. $X_{t_i} \rightarrow X_{t_j}, i \neq j$); \label{pt3}
  \item SNPs can depend on other SNPs (i.e. $X_{s_i} \rightarrow X_{s_j}, 
    i \neq j$). \label{pt4}
\end{enumerate}
We also assume that dependencies between traits broadly follow the temporal
order in which they are measured; for instance, traits that are measured when
a plant variety is harvested can depend on those that are measured while it is
still in the field (and obviously on the markers as well), but not vice versa.
In other words, Assumptions \ref{pt3} and \ref{pt4} define BNs that describe the
dependencies of phenotypes on genotypes in a \textit{prognostic} model, as 
opposed to a \textit{diagnostic} model in which genotypes depend on phenotypes.
The latter is often preferred over the former because it results in simpler
models when the $X_i$ are discrete \citep{pourret}; in that setting, the number
of parameters grows exponentially with the number of parents of each node. 
However, this is not the case here due to Assumptions \ref{pt1} and \ref{pt2}.
Under these assumptions, the local distribution $\Prob(X_{t_i} \given \Pi_{X_{t_i}})$
of each trait is a linear model of the form
\begin{align}
\label{eq:trait}
  X_{t_i} &= \boldsymbol{\mu}_{t_i} + \Pi_{X_{t_i}} \boldsymbol{\beta}_{t_i} + 
              \boldsymbol{\varepsilon}_{t_i} \\ \notag
      &= \boldsymbol{\mu}_{t_i} + \underbrace{X_{t_j} \beta_{t_j} + \ldots + 
          X_{t_k} \beta_{t_k}}_{\text{traits}} +
    \underbrace{X_{s_l} \beta_{s_l} + \ldots + X_{s_m} \beta_{s_m}}_{\text{SNPs}} +
      \,\boldsymbol{\varepsilon}_{t_i},&
    \boldsymbol{\varepsilon}_{t_i} \sim N(0, \sigma^2_{t_i}\mathbf{I})
\end{align}
where $\mathbf{I}$ is the identity matrix. SNPs will typically be coded using
their allele counts ($0, 1, 2$), although extensions to multiallelic SNPs and to
account for dominance are trivial. Similarly, the local distribution 
$\Prob(X_{s_i} \given \Pi_{X_{s_i}})$ of each SNP is
\begin{align}
\label{eq:snp}
  &X_{s_i} = \boldsymbol{\mu}_{s_i} + \underbrace{X_{s_l} \beta_{s_l} + \ldots + 
              X_{s_m} \beta_{s_m}}_{\text{SNPs}} + \,\boldsymbol{\varepsilon}_{s_i},&
  &\boldsymbol{\varepsilon}_{s_i} \sim N(0, \sigma^2_{s_i}\mathbf{I}).
\end{align}
Therefore, each parent only adds one parameter to a local distribution.

The regression parameters in (\ref{eq:trait}) and (\ref{eq:snp}) can be 
estimated in different ways. When $G$ is sparse, ordinary least squares (OLS)
are often used because each local distribution is estimated independently and
contains few regressors. Otherwise, penalized estimators such as ridge 
regression \citep[RR;][]{ridge} can be used when $G$ is dense. The resulting
BN can then be considered a flexible implementation of multivariate ridge
regression, which has a number of of desirable properties over OLS \citep{brown}.

Equivalently, we can describe a BN using its global distribution, denoted with
$\Prob(\mathbf{X})$ in (\ref{eq:markov}). Following Assumption \ref{pt1},
$\mathbf{X}$ has a multivariate normal distribution, say $\mathbf{X} \sim 
N(\boldsymbol{\mu}, \Sigma)$. In addition, by definition graphical separation
of two nodes $X_i$ and $X_j$ in $G$ implies the conditional independence of
the corresponding variables given the rest. As a result, some
elements of the precision matrix $\Omega = \Sigma^{-1}$ will be equal to zero
and some will be strictly positive according to the structure of $G$. The 
link with the parameterisation based on the local distributions arises from
the fact that in each $\Prob(X_{i} \given \Pi_{X_{i}})$ the regression 
coefficient associated with $X_j$ will be $\beta_j = - \Omega_{ij} /\Omega_{ii}$;
so $\beta_j = 0$ if and only if the $(i, j)$ element of $\Omega$ is itself 
equal to zero \citep[][pp. 68--69]{wermuth}.

It is interesting to note that this formulation defines BNs that are equivalent
to multivariate GBLUP models \citep{henderson}. For simplicity of notation,
assume we are modeling only two traits $X_{t_1}$ and $X_{t_2}$ with a common
set of SNP genotypes $\mathbf{X}_{\mathbf{S}}$. In this case a multivariate
GBLUP model has the form
\begin{equation}
\label{eq:multiblup}
  \left[
    \begin{array}{c}
      X_{t_1} \\
      X_{t_2}
    \end{array}
  \right]
  =
  \left[
    \begin{array}{c}
      \boldsymbol{\mu}_{t_1} \\
      \boldsymbol{\mu}_{t_2}
    \end{array}
  \right]
  +
  \left[
    \begin{array}{cc}
      \mathbf{Z}_{\mathbf{S}} & \mathbf{O} \\
      \mathbf{O} & \mathbf{Z}_{\mathbf{S}}
    \end{array}
  \right]
  \left[
    \begin{array}{c}
      \mathbf{u}_{t_1} \\
      \mathbf{u}_{t_2} 
    \end{array}
  \right]
  +
  \left[
    \begin{array}{c}
      \boldsymbol{\varepsilon}_{t_1} \\
      \boldsymbol{\varepsilon}_{t_2} 
    \end{array}
  \right]
\end{equation}
where $\mathbf{u}_{t_1}, \mathbf{u}_{t_2}$ are the random effects for the two
traits; $\mathbf{Z}_{\mathbf{S}}$ is the design matrix of the genotypes 
$\mathbf{X}_{\mathbf{S}}$; $\boldsymbol{\mu}_{t_1}, \boldsymbol{\mu}_{t_2}$
are the population means; and $\boldsymbol{\varepsilon}_{t_1}, 
\boldsymbol{\varepsilon}_{t_2}$ are the error terms. $\mathbf{u}_{t_1},
\mathbf{u}_{t_2}$ and $\boldsymbol{\varepsilon}_{t_1}, 
\boldsymbol{\varepsilon}_{t_2}$ are independent of each other and distributed
as multivariate normals with zero mean and covariance matrices
\begin{align}
\label{eq:mlmm}
  &\COV\left(\left[
    \begin{array}{c}
      \mathbf{u}_{t_1} \\
      \mathbf{u}_{t_2}
    \end{array}
  \right]\right) =
  \left[
    \begin{array}{cc}
      \mathbf{G}_{t_{1}t_{1}} & \mathbf{G}_{t_{1}t_{2}} \\
      \mathbf{G}_{t_{1}t_{2}}^T & \mathbf{G}_{t_{2}t_{2}} 
    \end{array}
  \right]&
  &\text{and}&
  &\COV\left(\left[
    \begin{array}{c}
      \boldsymbol{\varepsilon}_{t_1} \\
      \boldsymbol{\varepsilon}_{t_2} 
    \end{array}
  \right]\right) =
  \left[
    \begin{array}{cc}
      \sigma^2_{t_1} \mathbf{I} & \sigma^2_{t_1 t_2} \mathbf{I} \\
      \sigma^2_{t_2 t_1} \mathbf{I} & \sigma^2_{t_2} \mathbf{I}
    \end{array}
  \right].
\end{align}
The covariance matrix $\mathbf{G}_{t_{1}t_{2}}$ models the pleiotropic effects
of the SNPs on traits, potentially increasing the accuracy of multivariate GBLUP
compared to a single-trait model.
 
As was the case in (\ref{eq:trait}), each trait $X_{t_i}, i = 1,2$ has a 
population mean $\boldsymbol{\mu}_{t_i}$ and an error term 
$\boldsymbol{\varepsilon}_{t_i}$ that is normally distributed and independent
of the SNP effects. The residual variance $\sigma^2_{t_i}$ is also specific to
each trait. The two traits depend directly on each other because of the
covariances $\sigma^2_{t_1 t_2},\sigma^2_{t_2 t_1}$; and indirectly through
the covariance structure of the SNP effects $\mathbf{G}_{t_{1}t_{2}}$. If we
denote $\COV([\mathbf{u}_{t_1} \mathbf{u}_{t_2}]^T)$ as $\mathbf{G}$ and
$\COV([\boldsymbol{\varepsilon}_{t_1} \boldsymbol{\varepsilon}_{t_2}]^T)$
as $\mathbf{R}$, we can write
\begin{equation}
  \Sigma = 
  \COV\left(\left[
    \begin{array}{c}
      X_{t_1} \\
      X_{t_2} \\
      \hline
      \mathbf{u}_{t_1} \\
      \mathbf{u}_{t_2}
    \end{array}
  \right]\right)
  =
  \left[
   \begin{array}{c|c}
     \mathbf{Z}_{\mathbf{S}} \mathbf{G} \mathbf{Z}_{\mathbf{S}}^T + \mathbf{R} &
     \mathbf{Z}_{\mathbf{S}} \mathbf{G} \\
     \hline
     (\mathbf{Z}_{\mathbf{S}} \mathbf{G})^T &
     \mathbf{G}
   \end{array}
  \right]
\end{equation}
which is the covariance matrix of the global distribution. The structure of
the BN defined over $\mathbf{X} = \{X_{t_1}, X_{t_2}, \mathbf{u}_{t_1}, 
\mathbf{u}_{t_2}\}$ and corresponding to the multivariate GBLUP in 
(\ref{eq:multiblup}) arises from $\Omega = \Sigma^{-1}$ as discussed above.
Finally, it is important to note that even though GBLUP does not model the
SNP effects using the allele counts directly as in (\ref{eq:trait}) and
(\ref{eq:snp}), when $\mathbf{G}_{t_1,t_1}$ and $\mathbf{G}_{t_2,t_2}$ have
the form  $\mathbf{X_{\mathbf{S}}X_{\mathbf{S}}}^T$ the linear dependence on
$\mathbf{Z}_{\mathbf{S}}\mathbf{u}_{t_i}$ can be equivalently expressed 
as a random regression in the allele counts \citep{rrblup,rrblup2}.
The form of $\mathbf{G}_{t_1,t_1}, \mathbf{G}_{t_2,t_2}$ determines how the
allele counts are scaled or weighted in the regression. This formulation of
GBLUP results in a more natural interpretation of SNP effects, which is in
fact analogous to the interpretation they are given in a BN \citep{sagmb12}.

Another interesting property of the BN defined above is that the covariance
matrix of the SNP genotypes, which is a submatrix $\Sigma_{\mathbf{SS}}$ of 
$\Sigma$ (the global covariance matrix), is used in computing $\Omega$ and
determines which arcs are present in $G$ between the SNPs. Furthermore,
$\Sigma_{\mathbf{SS}}$ encodes the LD patterns between the SNPs as measured
by the squared allelic correlation $r^2$. This has been shown to be useful
in exploring complex LD patterns in an inbred Holstein cattle population,
albeit with a discrete BN \citep{gianola} and measuring LD in a way that is
closer to $D$ and $D'$ \citep{mackay}. Such patterns are reflected in the BN
through $\Omega$, providing an intuitive representation of LD as well as of 
genetic effects on phenotypes as a single, coherent whole.

BNs present two other advantages over classic multivariate regression models
such as multivariate GBLUP and ridge regression. Firstly, there is a vast 
literature on performing causal modeling with BNs from both experimental and
observational data \citep{causality}. Given the lack of a formal distinction
between response and explanatory variables in BNs, the same algorithms can be
used for inference on the traits based on the genotypes and vice versa. The
former includes the estimation of phenotypic EBVs, which is the basis of genomic
selection; the latter can be used for association mapping in polygenic traits
and when the desired phenotype is a combination of conditions on several
traits. Secondly, the fundamental properties of BNs do not depend on the
distributional assumptions of the data. Therefore, accommodating heterogeneous
traits (discrete, ordinal and continuous) in the model only requires to
specify the form of the local distributions.

Estimating a BN from data is typically performed as a two-step process. The 
first step consists in finding the graph $G$ that encodes the conditional
independencies present in the data, and is called \textit{structure learning}.
This can be achieved using conditional independence tests (\textit{constraint-based
learning}), goodness-of-fit scores (\textit{score-based learning}) or both
(\textit{hybrid learning}) to identify statistically significant arcs. The
second step is called \textit{parameter learning} and deals with the estimation
of the parameters of the local distributions; $G$ is known from the previous
step and defines which variables are included in each one. In addition, we
propose to use structure learning to retain in the BN only those SNPs that are
required to make inference on the traits and that make the remaining SNPs
redundant. For each trait, such a subset is called the \textit{Markov blanket}
\citep[$\mathcal{B}(X_{t_i})$;][]{pearl}, and includes the parents, the
children and the other nodes that share a child with the trait. Therefore, we
can disregard all the SNPs that are not part of any such Markov blanket and
reduce drastically the dimension of the model. We have shown in previous work
\citep{sagmb12} how Markov blankets are effective when used in this setting.

From these considerations, we used the R packages \textbf{bnlearn} \citep{jss09}
and \textbf{penalized} \citep{penalized} to implement the following hybrid approach
to BN learning.
\begin{enumerate}
  \item \textbf{Structure Learning.} 
    \begin{enumerate}
      \item For each trait $X_{t_i}$, use the SI-HITON-PC algorithm \citep{hitonpc}
        to learn the parents and the children of the trait; this is sufficient
        to identify $\mathcal{B}(X_{t_i})$ because the only nodes that can share a
        child with $X_{t_i}$ are other traits or SNPs that are parents of other
        traits due to Assumption \ref{pt3}. The choice of SI-HITON-PC is motivated
        by its similarity to single-SNP analysis, which is improved on with a 
        subsequent backward selection to remove false positives. Dependencies are
        assessed with Student's $t$-test for Pearson's correlation \citep{hotelling}
        and $\alpha = 0.01, 0.05, 0.10$.
      \item Drop all the markers which are not in any $\mathcal{B}(X_{t_i})$.
      \item Learn the structure of the BN from the nodes selected in the previous 
        step, setting the directions of the arcs according to the Assumptions 
        \ref{pt3} and \ref{pt4}. We identify the optimal structure as that 
        which maximizes the \textit{Bayesian information criterion} 
        \citep[BIC;][]{bic}.
    \end{enumerate}  
    \item \textbf{Parameter Learning.} Learn the parameters of the local 
      distributions using OLS and RR.
\end{enumerate}

For comparison, we also fitted an elastic net (ENET) model \citep{elastic} and
a univariate GBLUP individually on each trait and on all the available SNPs
using the \textbf{glmnet} \citep{glmnet} and \textbf{synbreed} \citep{synbreed}
R packages. Since we have shown BNs to be equivalent to a multivariate GBLUP,
we did not fit the latter as a separate model. We investigated the properties
of the resulting models using, in each case, $10$ runs of $10$-fold
cross-validation. Predictive power was assessed by averaging the cross-validated
correlations arising from the $10$ runs and computing confidence intervals as
in \citet{hooper}. In the case of BNs, predictions in the cross-validation
folds were performed jointly on all traits, and in two different ways: by
conditioning only on the SNPs in the BN, to provide a measure of
\textit{genetic predictive ability} ($\rho_G$) and a fair comparison with
single-trait models; and by conditioning on the parents of each trait, which
may in turn be traits themselves, to provide a tentative measure of
\textit{causal predictive ability} ($\rho_C$).

In order to perform inference, we produced an averaged BN using the $100$
networks we obtained in the course of cross-validation. First, we created an
averaged network structure using their graphs as in \citet{aime11}: we kept
only those arcs that appear with a frequency higher than a threshold estimated
from the graphs themselves. SNPs which ended up as isolated nodes (i.e. they
were not connected to any other SNP or trait) were dropped. We then estimated
the parameters of the averaged BN with RR using the whole data set. We used
the resulting BN to generate samples of $10^6$ random observations from the 
conditional distributions of various traits and SNPs with either logic
sampling or likelihood weighting \citep{koller}, in order to explore their
properties and interplay under different conditions. Statistics estimated from
such a big sample are very precise and can capture even small differences
reliably.

We based our analysis on a winter wheat population produced by the UK National
Institute of Agricultural Botany (NIAB) comprising  $15877$ SNPs for $720$
genotypes. Seven traits were measured: yield (YLD; t/ha), flowering time (FT;
$6-54$, aggregate of 5 scores taken at 3-7 day intervals), height (HT; cm), 
yellow rust in the glasshouse (YR.GLASS; $1-9$) and in the field (YR.FIELD; 
$1-9$), fusarium (FUS; $1-9$) and mildew (MIL; $1-9$). Disease scores from
$1$ to $9$ reflect increasing level of infection, and flowering time scores
from $6$ to $54$ increasing lateness in flowering. The population was created
using a Multiparent Advanced Generation Inter-Cross (MAGIC) scheme. Such a
scheme is designed to produce a mapping population from several generations
of intercrossing among $8$ founders, and has the potential to improve 
quantitative trait loci (QTL) mapping precision \citep[for more details
see][]{magic}. The use of multiple founder varieties results in a population
which is segregating for more QTLs and traits than a biparental population;
and the balanced crossing used in each generation reduces LD and family
structure by ensuring each founder has an equal opportunity to contribute to
each genotype.

SNPs were preprocessed by removing those with minor allele frequencies $< 1\%$
and those with $> 20\%$ missing data. Missing data in the remaining SNPs were
imputed using the \textbf{impute} R package \citep{impute}. Other widely used
imputation methods in genetics, such as that implemented in MaCH \citep{mach},
could not be used because of the lack of precise mapping information at the
time of the analysis; a 90K consensus map has just been submitted for 
publication \citep{90kmap}. Subsequently, we removed one SNP from each pair
whose allele counts have correlation $> 0.95$ to increase the numerical
stability of the models. In the end, $3164$ SNPs were left for analysis.
Phenotypes were adjusted for kinship using a univariate BLUP model for each
trait based on pedigree information, thus accounting for population structure.
Individuals with missing pedigree information or phenotypes were dropped from
the analysis, leaving $600$ individuals with complete records.

\section{Results}

\begin{table}
\begin{center}
\sf \footnotesize
  \begin{tabular}{lc|ccccccc}
                               &          & YLD & FT & HT & YR.FIELD & YR.GLASS & MIL & FUS \\
    \hline
    ENET                       & $\rho_G$                  & $0.15$
                                                           & $0.30$ 
                                                           & $0.48$
                                                           & $0.39$
                                                           & $0.59$
                                                           & $0.21$
                                                           & $0.27$ \\
    \hline
    GBLUP                      & $\rho_G$                  & $0.10$
                                                           & $0.15$
                                                           & $0.19$
                                                           & $0.22$
                                                           & $0.32$
                                                           & $0.21$
                                                           & $0.12$ \\
    \hline
    \multirow{2}{*}{BN,0.01}   & $\rho_G$                  & $0.20$
                                                           & $0.29$
                                                           & $0.46$
                                                           & $0.37$
                                                           & $0.60$
                                                           & $0.12$ 
                                                           & $0.22$ \\
                               & $\rho_C$                  & $0.38$
                                                           & $0.29$
                                                           & $0.45$
                                                           & $0.44$
                                                           & $0.62$
                                                           & $0.13$
                                                           & $0.33$ \\
    \hline
    \multirow{2}{*}{BN,0.05}   & $\rho_G$                  & $0.18$
                                                           & $0.27$
                                                           & $0.46$
                                                           & $0.39$
                                                           & $0.61$
                                                           & $0.12$
                                                           & $0.25$ \\
                               & $\rho_C$                  & $0.34$
                                                           & $0.27$
                                                           & $0.45$
                                                           & $0.44$
                                                           & $0.63$
                                                           & $0.14$
                                                           & $0.32$ \\
    \hline
    \multirow{2}{*}{BN,0.10}   & $\rho_G$                  & $0.18$ 
                                                           & $0.28$
                                                           & $0.45$
                                                           & $0.40$
                                                           & $0.62$
                                                           & $0.13$
                                                           & $0.25$ \\
                               & $\rho_C$                  & $0.34$
                                                           & $0.28$
                                                           & $0.45$
                                                           & $0.45$
                                                           & $0.63$
                                                           & $0.14$
                                                           & $0.31$ \\
    \hline
  \end{tabular}
  \caption{Genetic ($\rho_G$) and causal ($\rho_C$) predictive correlations for
    the $7$ traits and for single-trait elastic net (ENET), single-trait GBLUP
    and BNs estimated with $\alpha = 0.01, 0.05, 0.10$ and RR. Standard
    deviations computed as in \citet{hooper} is $0.01$ for all correlations.
    Traits are yield (YLD), flowering time (FT), height (HT), yellow rust in
    the field (YR.FIELD) and in the glasshouse (YR.GLASS), mildew (MIL), and
    fusarium (FUS).
}
  \label{tab:correlations}
\end{center}
\end{table}

Table \ref{tab:correlations} shows genetic predictive correlations ($\rho_G$)
and causal predictive correlations ($\rho_C$) for single-trait ENET, 
single-trait GBLUP and BNs fitted with $\alpha = 0.01, 0.05, 0.10$. Only the
results for BNs whose parameters are estimated with RR are reported, because
using OLS provides essentially the same performance. The average $\rho_G$ 
obtained with RR across all traits is $0.324$ for $\alpha = 0.01$, $0.327$ for
$\alpha = 0.05$ and $0.331$ for $\alpha = 0.10$, all with a standard deviation
of $\pm 0.004$; with OLS we obtain $0.322$ for $\alpha = 0.01$, $0.325$ for 
$\alpha = 0.05$ and $0.324$ for $\alpha = 0.10$, again with a standard deviation
of $\pm 0.004$. Similar considerations can be made for $\rho_C$.

First of all, we note that BNs and single-trait ENET have comparable predictive
power for $\rho_G$: BNs are best for YLD, YR.GLASS and YR.FIELD, while ENET is
best for FT, HT, MIL, and FUS. Overall, the average $\rho_G$ across all $7$ 
traits is $0.343 \pm 0.004$ for ENET and $0.331 \pm 0.004$ for BNs with 
$\alpha = 0.10$. Therefore, while ENET outperforms BNs on average, BNs still
provide the best $\rho_G$ in $3$ traits out of $7$. In addition, both ENET and
BNs outperform single-trait GBLUP, which has $\rho_G = 0.186 \pm 0.005$ overall.
As expected, the choice of the kinship matrix used in GBLUP does not significantly
affect $\rho_G$ because we accounted for the effect of family structure on the
traits as a preliminary step. Using different marker-based estimates of kinship
such as allele sharing \citep{habier} or allelic correlation \citep{astle}
provides no benefit over not using a kinship matrix at all.

It is also apparent that increasing $\alpha$ does not produce any marked
increase in $\rho_G$; while larger values of $\alpha$ result in larger BNs, 
the small increase in predictive power is not worth the longer time required
to estimate the model under cross-validation. On average, we learned BNs with
$47$ nodes (including the $7$ traits) in a few seconds for $\alpha = 0.01$;
with $75$ nodes in $20$ minutes for $\alpha = 0.05$; and with $89$ nodes
in $2.5$ hours for $\alpha = 0.10$. Further increasing $\alpha$ as in
\citet{sagmb12} only exacerbates the problem ($24$ days for $\alpha = 0.15$,
results not shown). Of all the SNPs included in BNs, few are not parents of any
trait and thus appear to be false positives: $1$ out of $40$ ($2.5\%$) for
$\alpha = 0.01$, $2$ out of $68$ ($2.9\%$) for $\alpha = 0.05$ and $4$ out of
$82$ ($4.8\%$) for $\alpha = 0.10$. The dimension of the BNs is in stark 
contrast with the average number of non-zero SNP effects in the ENET models:
$110$ non-zero coefficients for YR.GLASS, $2661$ for YLD, $55$ for HT, $105$
for YR.FIELD, $333$ for FUS, $1725$ for MIL and $24$ for FT. 

As far as causal predictive correlations $\rho_C$ are concerned, we observe a
distinct improvement compared to $\rho_G$ for $3$ traits: YLD, YR.FIELD and
FUS. As for the other $4$ traits, the difference between $\rho_G$ and $\rho_C$
is not as marked, even though it is statistically significant in all cases 
except flowering time. Overall, $\rho_C = 0.373 \pm 0.004$ which is higher than
both BN's $\rho_G = 0.331 \pm 0.04$ for $\alpha = 0.10$ and the ENET's 
$\rho_G = 0.343 \pm 0.004$.

\begin{figure}
\sf \footnotesize
  \includegraphics[width=\textwidth]{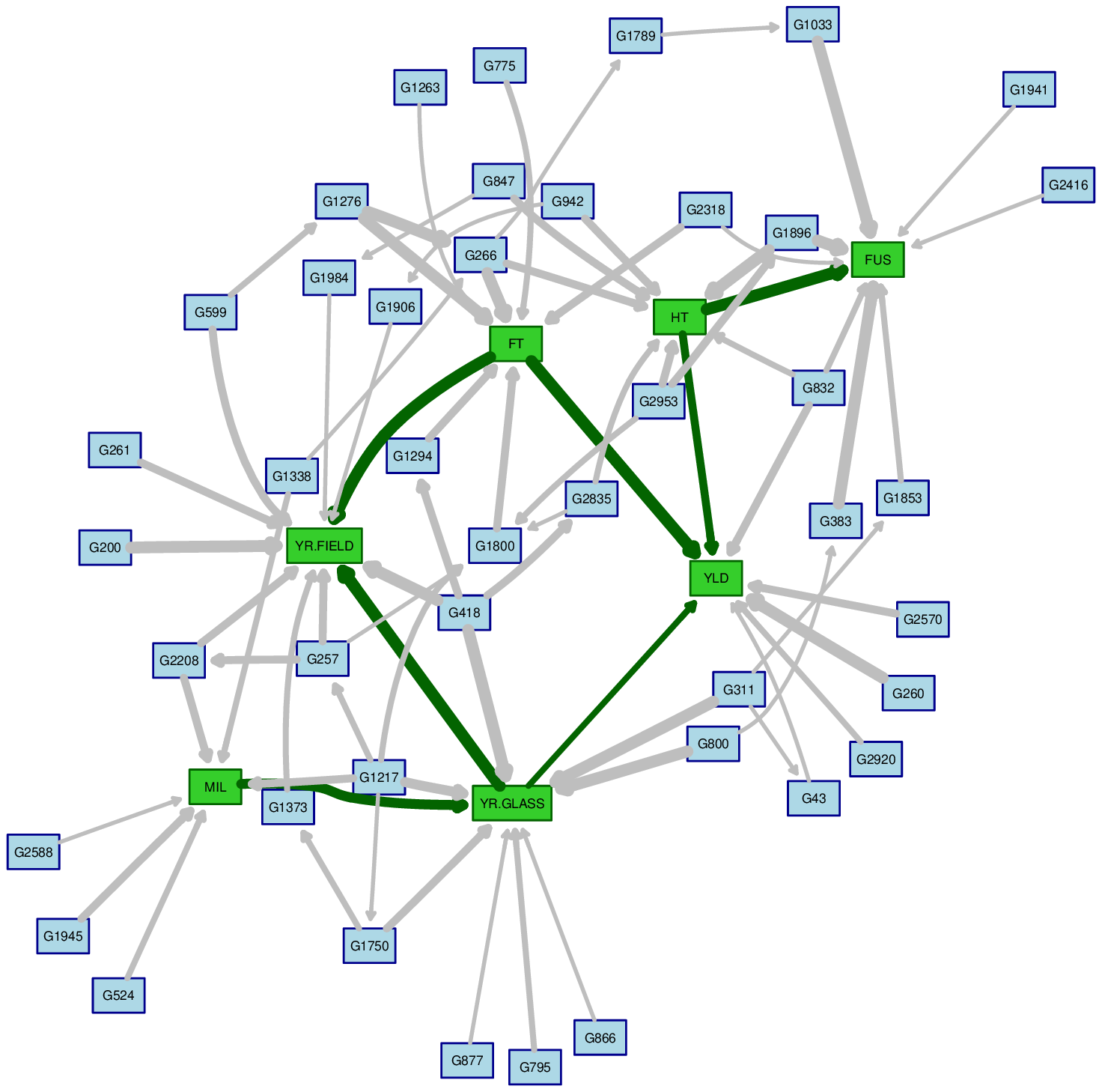}
  \caption{Averaged network obtained from the cross-validated BNs for
    $\alpha = 0.10$. Green nodes correspond to traits: yield (YLD), flowering
    time (FT), height (HT), yellow rust in the field (YR.FIELD) and in the 
    glasshouse (YR.GLASS), mildew (MIL), and fusarium (FUS). Blue nodes
    correspond to SNPs. The thickness of the arcs represents the strength of
    the corresponding dependence relationships as measured by their frequency
    in the BNs produced during cross-validation.}
  \label{fig:averaged}
\end{figure}

\begin{table}
\begin{center}
\sf \footnotesize
  \begin{tabular}{ll|ll}
    LABEL & NAME                     & LABEL & NAME                     \\
    \hline
    G418  & BobWhite\_c5756\_516     &  G311  & BobWhite\_c37358\_208   \\
    G800  & BS00022299\_51           &  G877  & BS00022830\_51          \\
    G866  & BS00022703\_51           &  G795  & BS00022270\_51          \\
    G2570 & Kukri\_c7241\_322        &  G260  & BobWhite\_c29014\_241   \\
    G832  & BS00022473\_51           &  G1896 & Excalibur\_c19078\_210  \\
    G2953 & Tdurum\_contig64772\_417 &  G942  & BS00024496\_51          \\
    G266  & BobWhite\_c30043\_150    &  G847  & BS00022562\_51          \\
    G2835 & RFL\_Contig4790\_1091    &  G200  & BobWhite\_c22728\_78    \\
    G2208 & IAAV1322                 &  G257  & BobWhite\_c28819\_733   \\
    G1906 & Excalibur\_c20837\_868   &  G261  & BobWhite\_c2905\_590    \\
    G1984 & Excalibur\_c37696\_192   &  G599  & BS00009575\_51          \\
    G383  & BobWhite\_c47401\_491    &  G2416 & Kukri\_c100613\_331     \\
    G1033 & BS00035141\_51           &  G1941 & Excalibur\_c27950\_459  \\
    G1853 & Excalibur\_c11795\_934   &  G1338 & BS00066211\_51          \\
    G524  & BS00000721\_51           &  G1945 & Excalibur\_c29304\_176  \\
    G1276 & BS00064538\_51           &  G1789 & D\_contig28346\_467     \\
    G2318 & IACX11305                &  G1800 & D\_GBUVHFX01DSLGX\_212  \\
    G1294 & BS00065110\_51           &  G775  & BS00022148\_51          \\
    G1750 & CAP12\_c2800\_262        &  G43   & BobWhite\_c11692\_148   \\
    G1373 & BS00067203\_51           &  G1217 & BS00062679\_51          \\
    G2588 & Kukri\_rep\_c102953\_304 &  G1263 & BS00064140\_51          \\
    G2920 & Tdurum\_contig42584\_1190\\
  \end{tabular}
  \caption{SNPs included in the averaged BN. The labels are those used in Figure
    \ref{fig:averaged}, while the SNP names are from \citet{magic} and
    \citet{90kmap}.}
  \label{tab:averaged}
\end{center}
\end{table}

The averaged BN for $\alpha = 0.10$ is shown in Figure \ref{fig:averaged}; it
has $50$ nodes and $78$ arcs. For ease of plotting, the SNP names corresponding
to the labels used in the figure are reported in Table \ref{tab:averaged}. The
dimension of the BN is comparable to that obtained for $\alpha = 0.01$ ($30$
nodes, $44$ arcs) and $\alpha = 0.05$ ($44$ nodes, $66$ arcs). In all three
cases the threshold for arc inclusion estimated as in \citet{aime11} is $0.49$,
which is close to the intuitive choice of including in the averaged BN those
arcs that appear in more than half of the BNs obtained during cross-validation.
All SNPs in the averaged BN are linked with at least one trait, with the 
exception of G1789 (D\_contig28346\_467). Their minor allele frequencies range
from $0.02$ (G2208; IAAV1322) to $0.47$ (G1945; Excalibur\_c29304\_176). 
Furthermore, the BN is small enough that RR and OLS parameter estimates are
practically equivalent.

As far as phenotypic traits are concerned, the averaged BN captures several
known relationships. YR.FIELD is influenced by FT (FT $\to$ YR.FIELD in Figure
\ref{fig:averaged}); early flowering genotypes will have their leaves exposed
to the pathogens for a longer time than later genotypes, resulting in higher
yellow rust scores even if they have the same level of true disease resistance.
This is substantiated by the posterior distribution of the disease score
conditional on flowering time being in the bottom quartile ($[21.0, 29.7]$) or
in the top quartile ($[33.8, 42.0]$): it has mean $2.54$ in the first case and
$2.33$ in the second. Standard deviation is $0.47$ in both cases. The same is
true for YR.GLASS, which has means $2.50$ and $2.48$ for early and late flowering
genotypes; standard deviation is $0.43$. The network structure suggests that
the YR.GLASS is not influenced directly by FT (\textit{i.e.} there is no FT
$\to$ YR.GLASS arc). The two yellow rust scores (YR.GLASS $\to$ YR.FIELD) are
positively correlated ($0.34$), likely because of durable resistance. In
addition, we note that YR.FIELD summarizes adult resistance to a mixed
population of pathotypes, which may include the specific pathotype used to
measure juvenile resistance in YR.GLASS.

We can also see from Figure \ref{fig:averaged} that YLD depends directly on
both HT (HT $\to$ YLD) and FT (FT $\to$ YLD); but it is affected only indirectly
by all the disease scores except YR.GLASS. Conditional on the combinations of
bottom and top quartiles for FT and HT ($[64.3, 74.5]$ and $[79.5, 87.7]$), the
expected yield is $7.54$, $7.71$, $7.15$ and $7.33$ respectively. Standard 
deviation is $0.47$ in all four scenarios. Therefore, we observe a marginal
increase in YLD of about $0.15$ when comparing short and tall genotypes, and a
marginal decrease of about $0.4$ when comparing early and late flowering 
genotypes; this is consistent with \citet{htyield} and \citet{ftyield}. The
interplay between HT and FT appears to be negligible in determining yield. 
Conditioning on the bottom and top quartiles of the disease scores, we see a
difference in the mean YLD of $+0.08$ (FUS), $-0.02$ (MIL), $-0.01$ (YR.GLASS)
and $-0.10$ (YR.FIELD).

The apparent increase in YLD associated with high FUS scores is the result of
the confounding effect of HT, which is directly linked to both variables in
the BN (FUS $\leftarrow$ HT $\rightarrow$ YLD). This is expected because 
susceptibility to fusarium is known to be positively related to HT 
\citep{fusarium}, which in turn affects YLD. Conditional on each quartile of
HT, FUS has a negative effect on YLD ranging from $-0.04$ to $-0.06$.

The last interaction between phenotypes in the BN is between MIL and YR.GLASS
(MIL $\to$ YR.GLASS). This can be explained by the increased susceptibility to
one disease in genotypes that are weakened by the onset of the other, by
disease resistance being controlled by shared regions in the genome 
\citep{mildewg1,mildewg2} and to a lesser extent by the influence of weather
conditions \citep{mildewp1}. The BN in Figure \ref{fig:averaged} identifies
$9$ SNPs that are linked to at least one of MIL and YR.GLASS, and may possibly
be tagging pleiotropic QTLs for disease resistance. By contrasting low and high
level of both diseases (scores $\leqslant 1.5$ and $\geqslant 3.5$, 
respectively), we can infer which allele may be linked with resistance to both
diseases using the conditional expected allele counts, $n_{\mathrm{LOW}}$ and
$n_{\mathrm{HIGH}}$. For $3$ of the $9$ genes the difference between the two
is marked: G418 (BobWhite\_c5756\_516; $n_{\mathrm{LOW}} = 0.5$, 
$n_{\mathrm{HIGH}} = 1.9$), G311 (BobWhite\_c37358\_208;
$n_{\mathrm{LOW}} = 1.1$, $n_{\mathrm{HIGH}} = 1.7$) and G1217
(BS00062679\_51; $n_{\mathrm{LOW}} = 0.8$, $n_{\mathrm{HIGH}} = 1.7$). 
The 90K consensus map in \citet{90kmap} locates G418 in chromosome 2D along with
other SNPs conferring resistance to YR.GLASS. The same is true also for G311 in
chromosome 2B, and for G2127 in chromosome 2A. As for the other $6$ SNPs,
$|n_{\mathrm{LOW}} - n_{\mathrm{HIGH}}| < 0.5$, which suggests that their
individual effects are small and that they might work in concert with other
genes producing polygenic effects.

Similar analyses on the other traits identify two more SNPs with
$|n_{\mathrm{LOW}} - n_{\mathrm{HIGH}}| \leqslant 0.5$ that may be tagging
known genes. G1896 (Excalibur\_c19078\_210) has $n_{\mathrm{LOW}} = 0.3$, 
$n_{\mathrm{HIGH}} = 1.2$ when contrasting top and bottom quartiles for HT;
and has $n_{\mathrm{LOW}} = 0.2$, $n_{\mathrm{HIGH}} = 1.7$ when contrasting
the bottom quartile of HT and $\mathrm{FUS} \geqslant 3.5$ with the top 
quartile of HT and $\mathrm{FUS} \leqslant 1.5$. The latter pair of scenarios
is motivated by the fact that taller plants are less susceptible to fusarium
than shorter plants. The LD analysis in \citet{magic} suggests that this SNP
is located in chromosome 4D in this population, and that it may be tagging
\textit{Rht-D1b}, a dwarfing gene which is also closely associated with
resistance to fusarium \citep{fusarium}. In addition, G266 (BobWhite\_c30043\_150)
appears to be located in chromosome 2D and to be tagging \textit{Ppd-D1},
which controls photoperiod response. Contrasting the bottom quartiles of both
FT and HT with the top quartiles we have $n_{\mathrm{HIGH}} = 0$ and
$n_{\mathrm{LOW}} = 0.8$.

\section{Discussion}

Modeling multiple quantitative traits simultaneously has been known to result
in better predictive power than targeting one trait at a time in the context
of additive genetic models \citep{henderson}. BNs provide a general framework
to estimate and analyze such models. They also provide an accompanying graphical
representation that is intuitive yet rigorous; a plot such as that in Figure
\ref{fig:averaged} can be very useful for exploratory analysis, to disseminate
results and to motivate further quantitative and qualitative analyses in
GWAS and GS studies.

From a theoretical point of view, BNs are more versatile than additive models
in common use. By assuming variables are normally distributed, we have shown
that BNs are in fact equivalent to multivariate GBLUP and, by extension of
single-trait GBLUP. Furthermore, the separation between structure and parameter
learning makes it possible to accommodate different parametric assumptions with
relatively few changes, and subsume models such as univariate and multivariate
ridge-regression \citep{ridge,brown}. As far as inference is concerned, several
established methods from the literature can be used to predict traits from
SNPs and vice versa; two examples are logic sampling and likelihood weighting 
\citep{koller}. Both allow to explore complex scenarios of practical relevance
by estimating informative statistics from the corresponding conditional 
distributions of traits and SNPs. This is made easier by the lack of a formal
distinction between response and explanatory variables in the BN, which is
central in traditional linear models. As a result, BNs can be used for 
association studies as well as genomic prediction. In the former, we can
condition on some complex combination of traits and predict the expected allele
counts of SNPs. Such an approach has the potential of detecting which SNPs tag
relevant QTLs and which of their alleles are favourable. In the latter, we have
shown that BNs are competitive with a state-of-the-art model such as
single-trait ENET when predicting traits from SNPs, and that they outperform
single-trait GBLUP for the population analysed in this paper. As evidenced by
the difference between $\rho_G$ and $\rho_C$, using BNs as a multi-trait model
and performing predictions based on those variables identified as putative causal
for each trait outperforms ENET as well by leveraging pleiotropic effects
\citep{hartley}. This shows it is possible to improve genomic selection for
traits that are expensive to measure by incorporating cheaper ones in the
predictions. Clearly, the impact of correlated phenotypes on the predictive
power of BNs depends on the strength of their correlation.

Based on the BN in Figure \ref{fig:averaged}, we can also observe some 
interesting properties of BNs as genetic models. Firstly, the difference in
the number of SNPs included in the BNs compared to the ENET models can be
attributed to the limited ability of BNs to capture small epistatic
effects \citep{epibn}. Consider, for instance, a polygenic effect in which
two SNPs are jointly associated with a trait but in which each SNP is not 
significant on its own. Such an effect will not be captured because both
SNPs will be discarded by the single-SNP screening performed at the beginning
of feature selection. As observed in other studies, this does not have a
significant impact on predictive ability if a large enough $\alpha$ threshold
is used, as Markov blankets are very effective at feature selection 
\citep{integrative,sagmb12}. Secondly, SNPs with pleiotropic effects are 
included in the BN even when association with a single phenotype is detected;
at that point they can be linked to all relevant phenotypes. This is the
case of the SNPs controlling resistance to both mildew and yellow rust 
discussed above. Furthermore, direct and indirect effects of such SNPs and
of traits are correctly separated for the observed traits, as in the case of
the fusarium effect on yield.

MAGIC populations provide an ideal starting point for fitting BNs. On the one
hand, the particular pattern of crosses used to produce a MAGIC population
results in a very low population structure. This reduces the confounding effect
of relatedness on the estimation of SNP effects \citep{astle} and on mapping
approaches based on LD \citep{magic}. On the other hand, the size of of the 
population is large enough to detect weak associations and associations with
rare variants. Both are in fact present in the averaged BN, which includes
SNPs with minor allele frequencies as low as $0.02$ and SNPs which are 
significant (e.g. for MIL and YR.GLASS) only when considering multiple traits
at the same time.

Finally, SNPs of interest can be made to segregate in the population by choosing
the founders appropriately, since balanced crosses ensure opportunities for
recombination among the founders. This is particularly important in modeling
multiple phenotypes, as we need to ensure as many relevant QTLs and genes
as possible are tagged to correctly dissect their genetic layout.

\section{Acknowledgments}

The work presented in this paper forms part of the MIDRIB project, which is
funded by the UK Technology Strategy Board (TSB) and Biotechnology \& 
Biological Sciences Research Council (BBSRC), grant TS/I002170/1. The MAGIC
population was developed within BBSRC Crop Science Initiative project 
BB/E007201/1. Field trials and SNP genotyping were funded by the NIAB Trust.

\end{document}